\documentclass[twocolumn,aps,prc,amsmath,amssymb,graphicx,longbibliography]{revtex4}

\usepackage{epsfig}
\usepackage{epsf}

\usepackage{bm}

\newcommand{\vm}{{\bf m}}

\newcommand{\Heff}{{\bf H}_{\rm eff}}

\begin{document}

\title{Geometric phase-control of a spin-torque oscillator}

\author{A. R\"uckriegel}
\affiliation{Institute for Theoretical Physics, Utrecht
	University, Princetonplein 5, 3584 CC Utrecht, The Netherlands}

\author{R. A. Duine}

\affiliation{Institute for Theoretical Physics, Utrecht
	University, Princetonplein 5, 3584 CC Utrecht, The Netherlands}

\affiliation{Department of Applied Physics, Eindhoven University of Technology, P.O. Box 513, 5600 MB Eindhoven, The Netherlands}

\date{\today{}}

\begin{abstract}
We show that the phase of a spin-torque oscillator generically acquires a geometric contribution upon slow and cyclic variation of the parameters that govern its dynamics. As an example, we compute the geometric phase that results from a cyclic excursion of the magnitude of the external field and the current. We consider thermal phase fluctuations and conclude that the geometric phase should be experimentally observable at room temperature with current setups. We briefly comment on arrays of spin-torque oscillators and possible applications. 
\end{abstract}


\maketitle

\section{Introduction}
\label{sec:intro}
Spin-torque oscillators are auto-oscillators, systems in which a time-independent drive results in self-sustained oscillations \cite{4802339,KIM2012217,7505988}. In spin-torque oscillators, periodic magnetization dynamics results from a steady injection of angular momentum, by means of a spin current, that overcomes relaxation. Their simplest implementation is based on the precession of a uni-axial single-domain magnet around its axis of symmetry. The spin-current is  then injected either by using an adjacent fixed ferromagnet to spin-polarize charge current, or by using the spin-Hall effect in an adjacent normal-metal layer. In the latter situation one also refers to spin-Hall oscillators.

Regardless of their precise implementation, spin-torque oscillators are interesting systems that exhibit a variety of non-linear physical phenomena, such as phase and frequency locking. Possible applications of spin-torque oscillators range from the emission and detection of microwave radiation, to neuromorphic computing \cite{Locatelli_2013}. Some of these applications rely on the control of the phase of the oscillator. An example of such an application is in the field of magnonics \cite{Kruglyak_2010}, where the phase of the oscillator could be imprinted on the phases of spin waves that perform certain logic operations by controlled spin-wave interference. Another example is that of associative memory applications that may be possible with the phase of spin-torque oscillators \cite{Locatelli_2013}.   

In this article we show that the phase of a spin-torque oscillator can be controlled geometrically. More precisely, we show that a sufficiently slow cyclic change of the parameters that govern the dynamics of the oscillator results in a geometric contribution to the phase shift. Here, geometric means that the phase shift is only determined by the geometry of the path in the parameter space, but, for example, not by how fast it is traversed --- provided the parameters are varied sufficiently slowly. As a concrete example, we consider the geometric phase that arises from a loop in the parameter space that is spanned by the magnitude of the external field and the magnitude of the spin current. 

Because spin-torque oscillators are dissipative systems, the geometric phase that is elucidated here is not straightforwardly related to well-known examples of geometric phases, such as the Berry phase in quantum mechanics \cite{berry1984quantal}, or the Hannay angle in classical mechanics \cite{Hannay_1985,RUCKRIEGEL2020168010}. The geometric phase we consider is rather an example of a geometric phase first pointed out by Landsberg \cite{PhysRevLett.69.865,Sinitsyn_2009} and by Ning and Haken \cite{PhysRevA.43.6410}. 

The remainder of this article is organized as follows: in the next section we go in detail through the specific example of the geometric phase that arises due to a cyclic variation of field and current. In Sec.~\ref{sec:thermalfluctuations} we discuss the effect of thermal phase fluctuations and conclude that the geometric phase should be observable despite these fluctuations. We end with a conclusion and outlook, where we also discuss possible extensions and applications of our work. In the appendices we consider a more general cyclic variation of control parameters, discuss the influence of ellipticity, and also provide some results for arrays of spin-torque oscillators. 

\begin{figure} 
	\includegraphics[width=10cm]{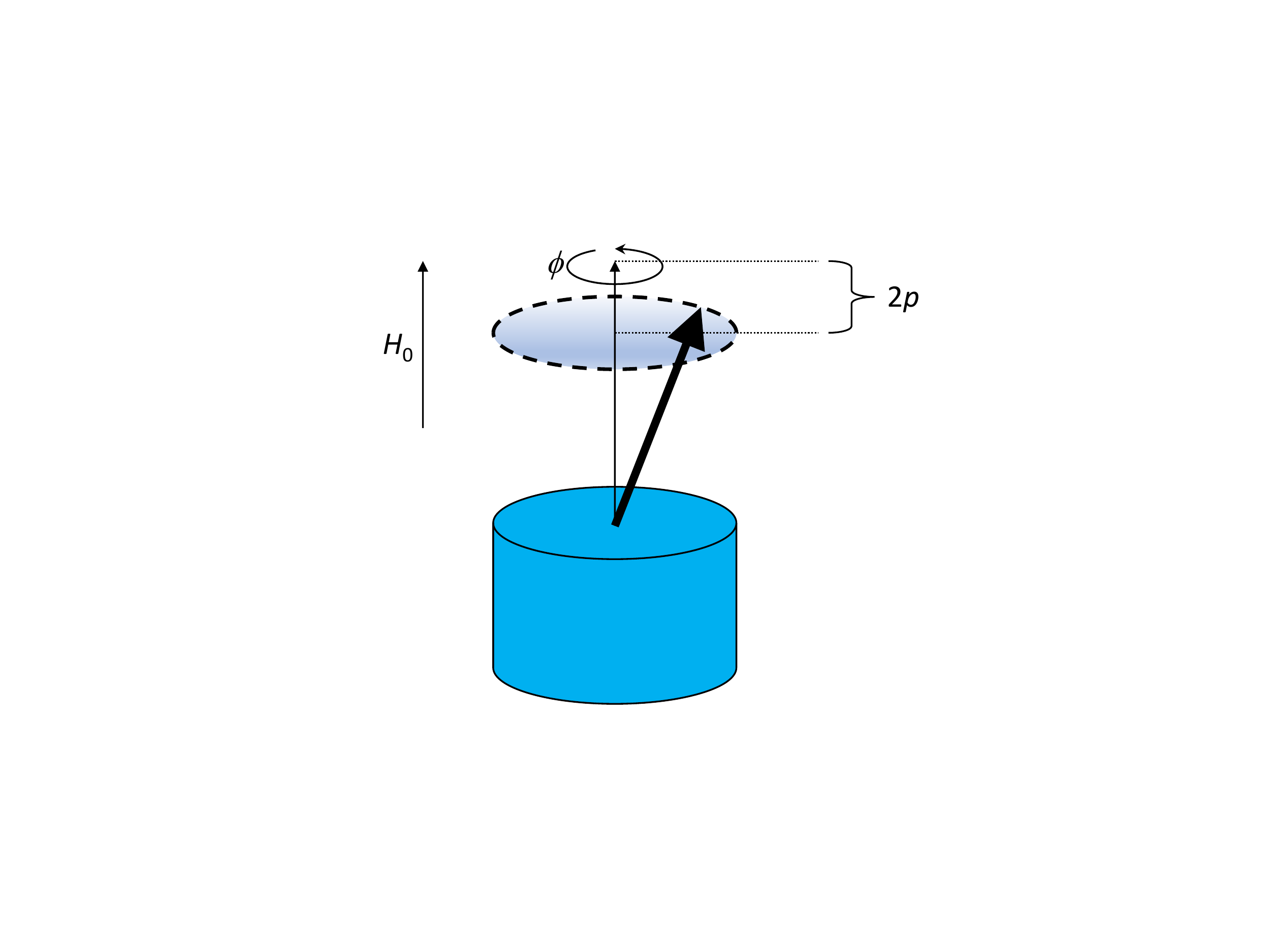} 
	\caption{Illustration of the set-up that is considered: a uni-axial single-domain magnet into which spin current (not shown) is injected. The external field with magnitude $H_0$ points in the $z$-direction. The magnetization dynamics is parametrized in terms of the azimuthal angle $\phi$ and the power $p$. The power determines the projection of the magnetization direction onto the $z$-axis.}
	\label{fig:setup}
\end{figure}

\section{Geometric phase due to cyclic variation of external field and current}
\label{sec:singledomain}
We consider a specific implementation of a spin-torque nano-oscillator based on a single-domain magnet with uni-axial symmetry into which spin current is injected (see Fig.~\ref{fig:setup}). Here, we mostly follow Slavin and Tiberkevich \cite{4802339} in the derivation of the equations of motion and the discussion of the critical current and the equilibrium power. We start from the Landau-Lifshitz-Gilbert equation for the magnetization direction $\vm$:
\begin{eqnarray}
\label{eq:LLG}
\frac{\partial \vm (t)}{\partial t} &=& - \gamma \mu_0 \vm (t) \times  \Heff (\vm (t))  \nonumber \\
&& - \alpha \gamma \mu_0\vm (t) \times  \left[\vm (t) \times  \Heff (\vm (t))\right] \nonumber \\
&& + \bm{\tau}_{\rm cit} (\vm (t))~,
\end{eqnarray}
where $\gamma>0$ is the modulus of the gyromagnetic ratio, $\mu_0$ is the vacuum permeability,  and $\Heff$ is the effective field. The Gilbert damping term is determined by the constant $\alpha \ll 1$. The current-induced torque $\bm{\tau}_{\rm cit} (\vm (t))$ could include a conventional spin-transfer torque, as well as a spin-orbit torque. 

The effective field consists of a demagnetizing field and an external field of magnitude $H_0$ in the $z$-direction, so that
\begin{equation}
\Heff (\vm) = (H_0-4 \pi M_s m_z) {\bf e}_z~,  
\end{equation}
where $M_s$ is the saturation magnetization and ${\bf e}_z$ the unit vector in the $z$-direction. 

We take the curent-induced torque of the form
\begin{equation}
\bm{\tau}_{\rm cit} = I_s g (\vm \cdot {\bf e}_z)\vm \times (\vm \times {\bf e}_z)~, 
\end{equation}
which physically corresponds to a spin current $I_s$ with spin polarization in the $z$-direction that is injected. The dimensionless function $g (\vm \cdot {\bf e}_z)$ is determined by the details of the set-up. 

\begin{figure} 
	\includegraphics[width=9cm]{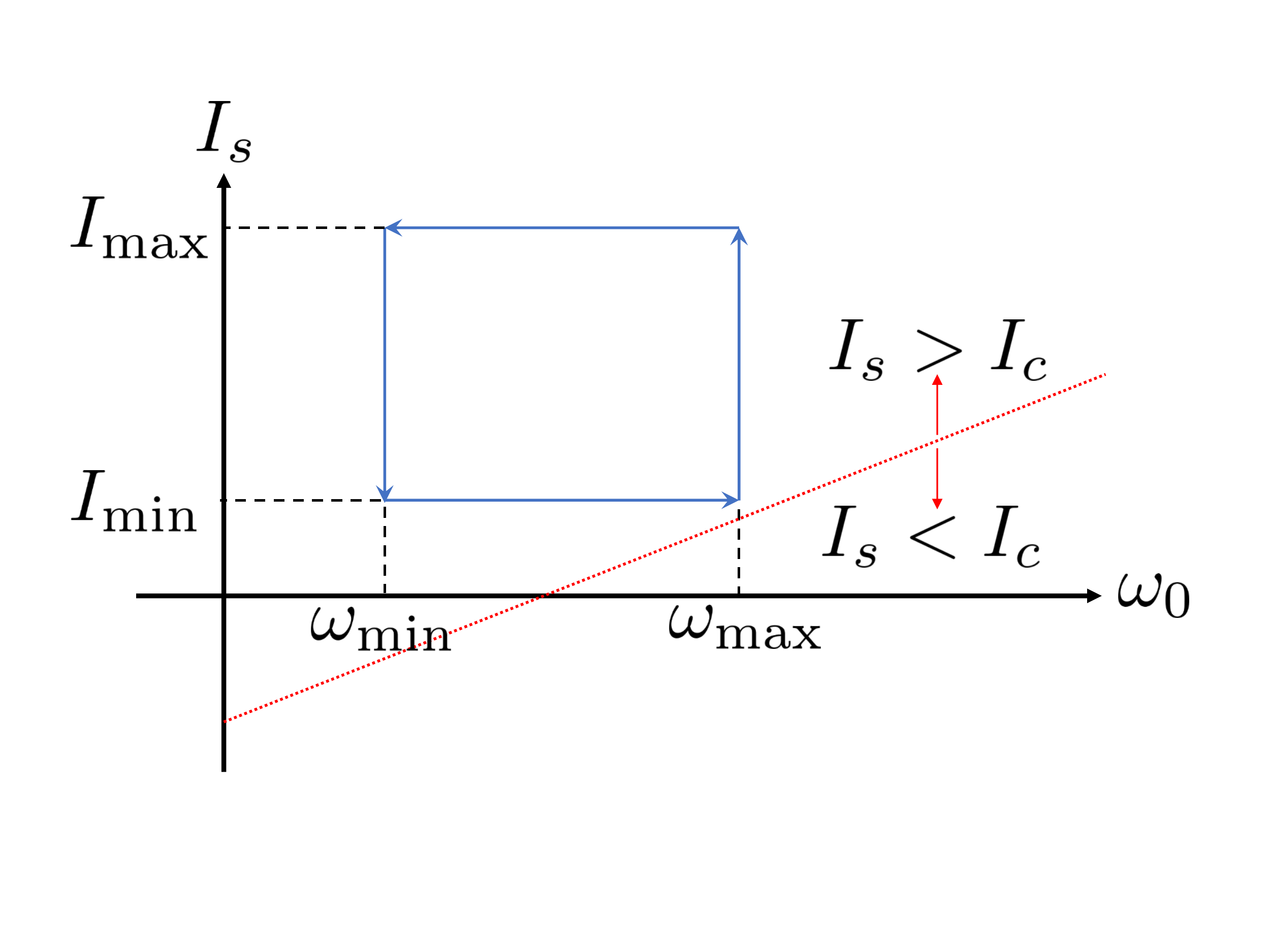} 
	\caption{Illustration of the rectangular loop in $(\omega_0,I_s)$-space that gives rise to the geometric phase. The dotted red line separates the subcritical ($I_s<I_c$) and supercritical ($I_s>I_c$) regions. }
	\label{fig:loop}
\end{figure}

We parametrize the magnetization direction by the power $0\le p(t) \le1$ and a precession angle $\phi (t)$ via
\begin{eqnarray}
\vm = \left(
\begin{array}{c}
2 \sqrt{p (1-p)} \sin \phi \\
2 \sqrt{p (1-p)} \cos \phi \\
1-2 p
\end{array}
\right)~.
\end{eqnarray}
This results in equations of the form 
\begin{subequations}
	\label{eq:EOMphiandpsingledomain}
	\begin{eqnarray}
	\dot p(t) &=& \!\!\! -2  \left[ \Gamma_+ (p (t);\omega_0,I_s) \!-\! \Gamma_- (p (t);\omega_0,I_s)\right] p (t),
	\\
	\dot \phi (t) &=& \omega (p (t);\omega_0,I_s),
	\end{eqnarray}
\end{subequations}
 with 
\begin{subequations}
	\label{eq:damping}
	\begin{eqnarray}
	\Gamma_+ (p;\omega_0,I_s) &=& \alpha \left(\omega_0 - \omega_M \right) - \alpha (\omega_0-3 \omega_M) p  \label{eq:dampingg}~,
	\\
	\Gamma_- (p;\omega_0,I_s) &=& I_s \left[\eta  + \left(\eta' - \eta\right) p\right]~, \label{eq:antidampingg}
	\end{eqnarray}
\end{subequations}
where $\eta (p) = g (1-2p)$, $\eta=\eta (0)$, and $\eta'=\left.d\eta (p)/dp\right|_{p=0}$. The precession frequency is given by
\begin{equation}
\label{eq:frequencystno}
\omega (p;\omega_0,I_s) = \omega_0-\omega_M +2 \omega_M p~.
\end{equation}
In deriving the above, we have kept terms up to quadratic order in $p$, and defined $\omega_0 = \gamma \mu_0 H_0$ and $\omega_M=4\pi \gamma \mu_0 M_s$. We also note that, while the coefficients in Eqs.~(\ref{eq:damping}) and (\ref{eq:frequencystno}) depend on the specific implementation, the various terms that arise are generic and the results derived below can be easily adopted to other implementations of spin-torque oscillators, such as vortex oscillators.

From the equations of motion we find that the critical spin current, above which the power $p$ becomes nonzero, is determined by $\Gamma_+ (0;\omega_0,I_s)=\Gamma_- (0;\omega_0,I_s)$, which yields
\begin{equation}
I_{\rm c} = \frac{\alpha}{\eta} \left( \omega_0 - \omega_M \right)~,
\end{equation}
The stationary power, found by solving for $p$ in the equation $\partial p/\partial t=0$, is 
\begin{equation}
p_0 (\omega_0,I_s) = \frac{\eta I_s - \alpha (\omega_0-\omega_M)}{(\eta-\eta') I_s+\alpha (3 \omega_M-\omega_0)}~.
\end{equation}

We are now in the position to compute the geometric phase that arises from a slow variation of the current and the external field, in such a way that they map out a closed loop in $(\omega_0,I_s)$-space (see Fig.~\ref{fig:loop}). Here, we adapt the discussion of Ref.~\cite{Sinitsyn_2009} to our specific case.
For  slowly-varying current $I_s (t)$ and field $\omega_0 (t)$, we have
\begin{equation}
\label{eq:poft}
p (t) = p_0 (t) + \left(\left. \frac{\partial  F }{\partial p} \right|_{p_0}\right)^{-1} \frac{\partial  p_0 (t)}{\partial t}~,
\end{equation}
where
\begin{equation}
F (p) = -2  \left[ \Gamma_+ (p;\omega_0,I_s)- \Gamma_- (p;\omega_0,I_s)\right] p~,
\end{equation}
and $p_0 (t)= p_0 (\omega_0 (t),I_s (t))$. Note that for stability of the auto-oscillations we should have that $\left. \partial  F /\partial p \right|_{p_0}<0$. Inserting the result for $p(t)$ in the equation for $\dot \phi (t)$, and expanding to linear order in $\partial p_0/\partial t$, we find that
\begin{eqnarray}
\label{eq:phidotadiabaticsingledomain}
\dot \phi (t) &=& \omega (p_0(t);\omega_0 (t),I_s (t)) \nonumber \\
&& + \left(\left. \frac{\partial \omega }{\partial p} \right|_{p_0} \right)  \left(\left. \frac{\partial  F }{\partial p} \right|_{p_0}\right)^{-1} \frac{\partial  p_0 (t)}{\partial t}~.
\end{eqnarray}
For closed loops in parameter space, starting, e.g., at $t=0$ and ending at $t=T$, integration of Eq.~(\ref{eq:phidotadiabaticsingledomain}) gives two contributions. The first, $\int_0^T \omega (p_0 (t);\omega_0(t),I_s (t)) dt$ is the dynamic phase; the second contribution is the geometric phase that we are after. It is given by
\begin{equation}
\label{eq:phigeoinit}
\phi_{\rm geo} =  \int_0^T dt\left[\left(\left. \frac{\partial \omega }{\partial p} \right|_{p_0} \right)  \left(\left. \frac{\partial  F }{\partial p} \right|_{p_0}\right)^{-1} \frac{\partial  p_0 (t)}{\partial t} \right]~.
\end{equation}
The above phase shift is dubbed a geometric phase shift as it does not depend on the specific path $(\omega_0 (t), I_s (t))$ that is traversed, i.e., it does not depend on the time-dependence of $\omega_0 (t)$ and $I_s (t)$, but only on the geometry of the path. To see this explicitly, we use that Eq.~(\ref{eq:phigeoinit}) is the parametrization of a line integral. The geometric phase can therefore be written as 
\begin{equation}
\label{eq:geophasevectorpotspec}
\phi_{\rm geo} =  \oint d\bm{\lambda}\cdot {\bf A}~, 
\end{equation}
with the vector potential
\begin{equation}
\label{eq:vectorpotspec}
A_i = \left(\left. \frac{\partial \omega }{\partial p} \right|_{p_0} \right)  \left(\left. \frac{\partial  F }{\partial p} \right|_{p_0}\right)^{-1} \frac{\partial  p_0 }{\partial \lambda_i}~,
\end{equation} 
where $i \in \{1, 2\}$, and $\bm{\lambda} = (\omega_0,I_s)^T$, so that $\lambda_1=\omega_0$ and $\lambda_2=I_s$. This vector potential is straightforwardly evaluated, and, with the help of Stokes' theorem, we rewrite the line integral for the geometric phase in Eq.~(\ref{eq:geophasevectorpotspec}) in terms of the rotation of the vector potential. This yields
\begin{equation}
\label{eq:geophasesingledomain}
\phi_{\rm geo} = \int_{O} d\omega_0 dI_s B (\omega_0,I_s)~,
\end{equation}
with 
\begin{eqnarray}
\label{eq:analyticexprgeophasecirc}
&&B (\omega_0,I_s) \equiv \frac{\partial A_{I_s}}{\partial \omega_0} - \frac{\partial A_{\omega_0}}{\partial I_s}\nonumber \\
&&=\frac{\alpha \eta'}
{\left[ \eta I_s - \alpha (\omega_0-\omega_M)\right]
	\left[(\eta-\eta') I_s+\alpha (3 \omega_M-\omega_0)\right]^2
}~, \nonumber \\
\end{eqnarray}
and where the integration is over the area enclosed by the loop in $(\omega_0,I_s)$-space. Note that this result shows that the geometric phase is only nonzero when $\eta'\neq 0$. For spin currents that approach the threshold current from above, i.e., for $I_s \downarrow I_c$, we have that $B (\omega_0,I_s)$ diverges as $B (\omega_0,I_s) \propto 1/(I_s-I_c)$. In this article we do not explore the dependence of the geometric phase on the various parameters of the system in detail but instead discuss an example in what follows. We note, however, that the geometric phase is in general nonzero, and, depending on parameters, can take any value between $0$ and $2\pi$. 

To illustrate the above result, we consider for simplicity a rectangular loop in parameter space, as illustrated in Fig.~\ref{fig:loop}. That is, starting from the initial values $\omega_0=\omega_{\rm min}$ and $I_s=I_{\rm min}$, we first increase the field linearly in time to $\omega_0=\omega_{\rm max}$. Keeping the field at this value, the spin current is increased linearly in time from $I_s=I_{\rm min}$ to $I_s=I_{\rm max}$. Hereafter, the field is decreased linearly in time back to its initial value $\omega_0=\omega_{\rm min}$, followed by decreasing the spin current back to $I_s=I_{\rm min}$. For all instantaneous values of spin current and field we should have that $I_s > I_c$. 

\begin{figure} 
	\includegraphics[width=8.5cm]{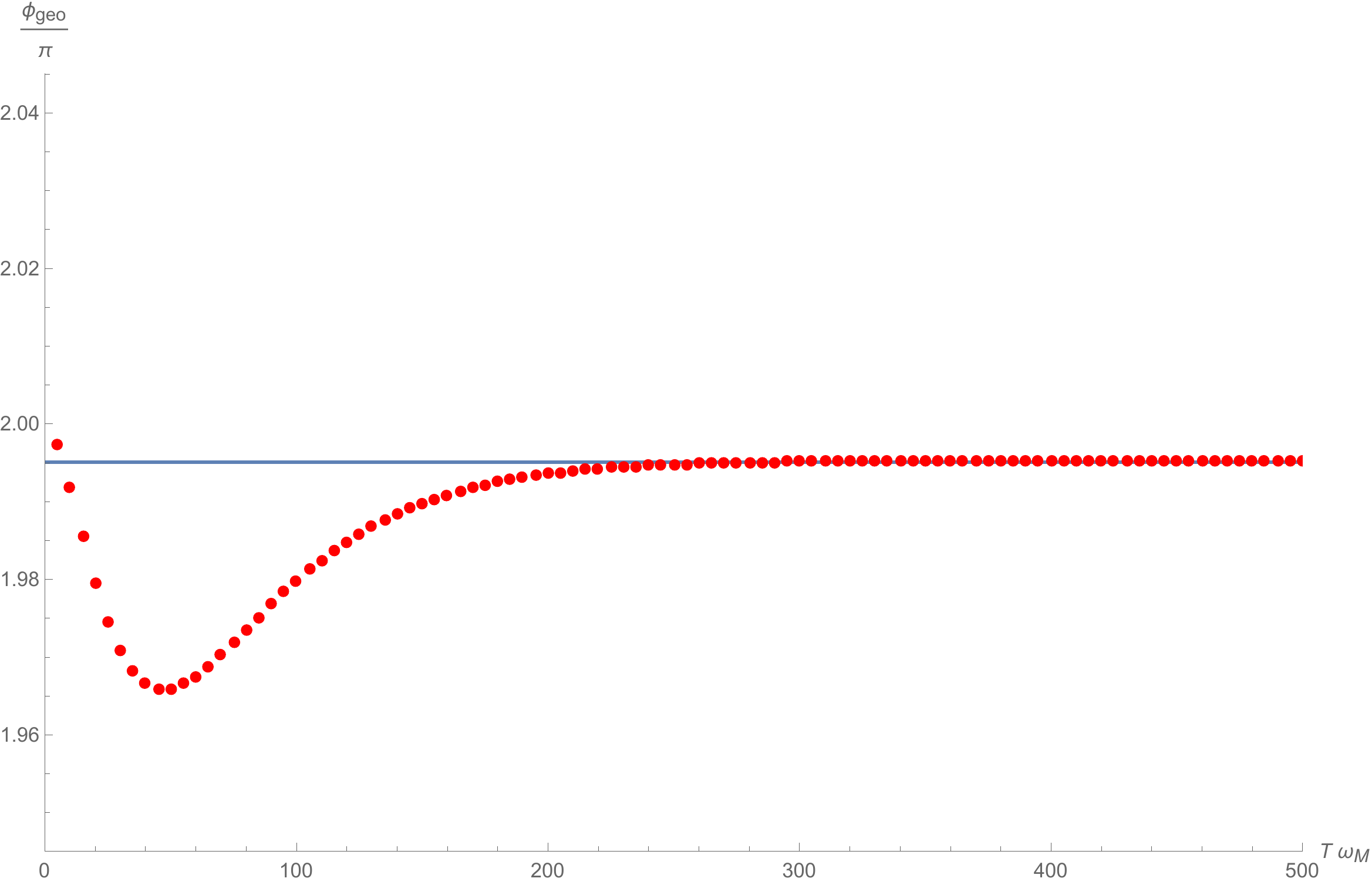} 
	\caption{Geometric phase for a rectangular loop in the parameter space spanned by external field and spin current as a function of total time $T$. Dots: numerical results, solid line: analytic results. Parameters taken are $\omega_{\rm min}/\omega_M=3/2$, $\omega_{\rm max}/\omega_M=2$, $\alpha=0.02$, $I_{\rm min}/\omega_M=0.061$,  $I_{\rm max}/\omega_M=0.066$, $\eta=1$, and $\eta'=0.1$.}
	\label{fig:geophasecirc}
\end{figure}

The numerical result for the geometric phase is shown in Fig.~\ref{fig:geophasecirc} as a function of the time $T$. This time is the total time over which the equation of motion~(\ref{eq:EOMphiandpsingledomain}) with Eqs.~(\ref{eq:damping}) are solved numerically. The field and spin current are kept constant for a time $T/6$, after which the four steps in the loop described above are performed for a time $T/6$ each. After this, the systems is evolved for constant field and current for a time $T/6$. To numerically determine the geometric phase, we have performed the loop both clockwise and counter-clockwise, and taken the difference of the phases after the loop and divided it by two. This cancels the dynamical phase. This result is then taken ${\rm mod}~2 \pi$. The analytic result in Eq.~(\ref{eq:analyticexprgeophasecirc}) is evaluated and yields a rather lengthy expression that we omit here. It is also plotted in Fig.~\ref{fig:geophasecirc}. From this figure, one sees that the numerical result for the phase approaches the analytic one for large times $\omega_M T \gg 1$. As the analytic result assumes the adiabatic limit, we conclude that the adiabatic limit is obtained when $\omega_M T \gg 1$. This is expected because it physically corresponds to many cycles of precession while the field and spin current are slowly varied. Contrary to the geometric phase which saturates for sufficiently large times, the dynamical phase (not plotted) increases linearly with time for large times $T$, as expected. 

\section{Thermal fluctuations}
\label{sec:thermalfluctuations}
We now investigate the effect of thermal fluctuations that may randomize the phase and render the geometric phase unobservable for long times. The starting point is the equations for power and phase that include thermal fluctuations via Langevin forces \cite{4802339,KIM2012217}:
\begin{subequations}
	\label{eq:EOMphiandplangevin}
	\begin{eqnarray}
	\dot p(t) &=& \!\!\!-2  \left[ \Gamma_+ (p (t)) \!- \!\Gamma_- (p(t))\right] p (t)  \nonumber \\
	&&+ 2 \sqrt{p (t)} \eta_p (t),
	\\
	\dot \phi (t) &=& \!\!\! \omega (p (t)) + \frac{\eta_\phi (t)}{\sqrt{p (t)}},
	\end{eqnarray}
\end{subequations}
where we suppressed, in the notation, the dependence of $\Gamma_-$, $\Gamma_+$, and $\omega$ on field and current. Here, the Langevin forces are mutually uncorrelated, have zero mean, and autocorrelations given by
\begin{equation}
\langle \eta_p (t) \eta_p (t') \rangle = \langle \eta_\phi (t) \eta_\phi (t') \rangle = \frac{\gamma \Gamma_+ (p(t))}{\beta M_s V_{\rm eff}\omega (p(t)) } \delta (t-t')~,
\end{equation}
in which $\langle \cdots \rangle$ denotes averaging over noise realizations, and where $\beta$ is the inverse thermal energy and $V_{\rm eff}$ is effective magnetic volume of the oscillator. The above Langevin equations for the power and phase of the oscillator are derived in the limit $p \ll 1$ from the stochastic Landau-Lifshitz-Gilbert equation \cite{PhysRev.130.1677}. In doing so, the classical limit is assumed, i.e., $\beta \hbar \omega \ll 1$. We write [see Eq.~(\ref{eq:poft})]
\begin{equation}
p (t) = p_0 (t) + \left(\left. \frac{\partial  F }{\partial p} \right|_{p_0}\right)^{-1} \frac{\partial  p_0 (t)}{\partial t} + \delta p (t)~,
\end{equation}
where $\delta p = {\mathcal O} \left(\eta_p\right)$. To be able to follow the developments in Sec.~4 of Ref.~\cite{KIM2012217}, we assume that we are sufficiently far in the supercritical region to take $\delta p$ small, and consider only the first order in $\partial  p_0 (t)/\partial t$ and $\delta p (t)$. This yields
\begin{equation}
\delta \dot p (t) =  \left(\left. \frac{\partial  F }{\partial p} \right|_{p_0}\right) \delta p + 2\sqrt{p_0} \eta_p (t)~.
\end{equation}
We assume an initial state at $t=0$ with $\delta p (0)=0$ and a well-defined phase. The equation for $\delta p (t)$ is then solved by
\begin{eqnarray}
&& \delta p (t) =  e^{\int_{0}^{t} \left(\left. \frac{\partial  F }{\partial p} \right|_{p_0}\right) dt'} 
\nonumber \\
&& \times \int_{0}^{t}
dt'' 2\sqrt{p_0} \eta_p (t'')  e^{-\int_{0}^{t''} \left(\left. \frac{\partial  F }{\partial p} \right|_{p_0}\right) dt'''}
~.
\end{eqnarray}
We replace the integrals in the exponents in the above by $ \left(\left. \partial  F /\partial p \right|_{p_0}\right) t$, and evaluate $p_0$ at time $t$ everywhere. This causes an error $\mathcal O \left(\partial  p_0 (t)/\partial t\right)$ and can be neglected in the above formal solutions since it gives rise to terms $\mathcal O \left( \eta_p \partial  p_0 (t)/\partial t  \right)$, which we ignored from the outset. This yields
\begin{equation}
\delta p (t) = 2\sqrt{p_0} e^{\left(\left. \frac{\partial  F }{\partial p} \right|_{p_0}\right) t} \int_{0}^{t}
dt' \eta_p (t')  e^{-\left(\left. \frac{\partial  F }{\partial p} \right|_{p_0}\right) t'}
~, 
\end{equation}
from which we find
\begin{eqnarray}
&& \langle \delta p (t) \delta p (t') \rangle = 
- \frac{2 \gamma p_0 \Gamma_+ (p_0)}{\beta M_s V_{\rm eff} \omega (p_0) \left(\left. \frac{\partial  F }{\partial p} \right|_{p_0}\right)} \nonumber\\
&& \times \left[
e^{\left(\left. \frac{\partial  F }{\partial p} \right|_{p_0}\right) |t-t'|}
- e^{\left(\left. \frac{\partial  F }{\partial p} \right|_{p_0}\right) (t+t')}
\right]~,
\end{eqnarray}
where $p_0$ is taken at time $t$. With this, we can evaluate the variance in the phase after a time $T$, given by 
\begin{eqnarray}
\left\langle\left(\Delta \phi (T)\right)^2 \right\rangle 
&=&\int_0^T dt \int_0^T dt' \left(\left. \frac{\partial \omega }{\partial p} \right|_{p_0} \right)^2 \langle \delta p (t) \delta p (t') \rangle \nonumber \\
&+& \int_0^T dt \int_0^T dt' \frac{1}{p_0} \langle \eta_\phi (t) \eta_\phi (t') \rangle~,
\end{eqnarray}
yielding
\begin{widetext}
\begin{equation}
\left\langle\left(\Delta \phi (T)\right)^2 \right\rangle 
= \frac{\gamma \Gamma_+ (p_0)}{ \beta M_s V_{\rm eff} \omega (p_0) }  
 \left\{\frac{T}{p_0}  +  p_0 \left(\frac{2 \left. \frac{\partial \omega }{\partial p} \right|_{p_0} }{\left. \frac{\partial  F }{\partial p} \right|_{p_0}}  \right)^2
\left[ T 
+ 
\frac{\left(1-
	e^{\left(\left. \frac{\partial  F }{\partial p} \right|_{p_0}\right) T} \right)
	\left(3-
	e^{\left(\left. \frac{\partial  F }{\partial p} \right|_{p_0}\right) T} \right)}
	{2\left(\left. \frac{\partial  F }{\partial p} \right|_{p_0}\right)}\right]\right\}~,
\end{equation}
\end{widetext}
where we take $p_0=p_0 (T)$.
In the above we have ignored the time dependence of $p_0$ in carrying out the various integrations, cf. our earlier approximations. 
To investigate the relative importance of phase fluctuations, we take the same parameters as in Fig.~\ref{fig:geophasecirc}, i.e.,  $\omega_{0}/\omega_M=3/2$, $\alpha=0.02$, $I_s/\omega_M=0.061$, $\eta=1$, and $\eta'=0.1$. We find that for these parameters and for $\omega_M T \gg 1$ the fluctuations in the phase are dominated by the second term in the above, which is estimated as
\begin{equation}
\left\langle\left(\Delta \phi (T)\right)^2 \right\rangle 
\sim \frac{\gamma T}{ \alpha \beta M_s V_{\rm eff} }~.
\end{equation}
That this term dominates for $\omega_M T \gg 1$ is understood as it is a factor $1/\alpha^2 \gg 1$ larger than the other term linear in $T$, whereas the other terms are either constant with $T$
or exponentially suppressed. The adiabatic regime where the geometric phase manifests was found to be reached when $\omega_M T \sim 100$. We demand that $\sqrt{\left\langle\left(\Delta \phi (T)\right)^2 \right\rangle }$ is at least one order of magnitude smaller than $\phi_{\rm geo}=\mathcal{O }\left(1\right)$. Taking $\beta \hbar \omega_M \sim 10^{-2}$, for a typical frequency of $\omega_M = 1$ GHz and room temperature, one requires that $\alpha M_s V_{\rm eff}/\gamma \hbar$ is at least $10^6$ to observe the geometric phase. The factor $M_s V_{\rm eff}/\gamma \hbar$ is the effective number of spins in the oscillator. For typical Gilbert damping $\alpha \sim 0.01$, one needs $M_s V_{\rm eff}/\gamma \hbar \sim 10^8$, which should be achievable. We conclude that observation of the geometric phase should be experimentally possible at room temperature, despite the thermal fluctuations. To reduce the effect of fluctuations and increase phase-stability, one may also consider arrays of spin-torque oscillators. In the appendix we show that such arrays exhibit a similar geometric phase. 

\section{Conclusions, discussion, and outlook}
\label{sec:concl} In conclusion, we have shown that the phase of a spin-torque oscillator picks up a geometric contribution when the parameters that govern its dynamics perform a loop in parameter space. We have focused on a spin-torque oscillator based on a uni-axial single-domain magnet and considered the geometric phase due to a cyclic excursion of the magnitude of field and current. In the appendices we consider the generic case and give a general expression for the geometric phase. There, we also show that the global phase of an array of coupled spin-torque oscillators acquires a similar geometric contribution. 

As the phase of a spin-torque oscillator can be measured directly (see e.g. Ref.~\cite{doi:10.1063/1.3467043}), our findings could be tested experimentally in a straightforward manner. In particular, our estimates indicate that, even at room temperature, thermal fluctuations do not render the geometric phase unobservable. To experimentally extract the geometric phase one could start from a state with well-defined phase by locking the phase of the oscillator to an external alternating source, and then performing a loop in the parameter space of field and current. By repeating the experiment, but with a reversed loop, and taking the difference of the phase between the forward and reversed loop, one would cancel the dynamical phase and directly obtain the geometric one times a factor of two. 

The geometric phase that we considered here could, for example, be used to imprint, in a controlled and reproducible way, a phase difference on two identical spin-torque oscillators that are initially phase locked. Such a phase difference could then be transferred to spin waves in setups where the oscillators act as spin-wave emitters, and could be useful for magnonic operations that rely on phase control of the spin waves \cite{Kruglyak_2010}. Other possible applications may be found in the context of neuromorphic computing based on spin-torque oscillators in which the phase plays an important role \cite{Locatelli_2013}.

Possible extensions of our work could be done in the direction of spin-torque oscillators based on antiferromagnets, or could be geared towards specific experimental implementations. We hope that this work stimulates efforts in these directions. 

\acknowledgements

This work is supported by the European Research Council via Consolidator Grant number
725509 SPINBEYOND. RD is member of the D-ITP consortium, a program of the Netherlands Organisation for
Scientific Research (NWO) that is funded by the Dutch
Ministry of Education, Culture and Science (OCW). This research was supported in part by the National Science Foundation under Grant No. NSF PHY-1748958.

\appendix

\section{General case}
\label{sec:geophase} 
The general equations of motion for the power $p (t)$ and phase $\phi (t)$ of a spin-torque nano-oscillator are given by 
\begin{subequations}
	\label{eq:EOMphiandp}
	\begin{eqnarray}
	\dot p(t) &=& -2  \left[ \Gamma_+ (p;\bm{\lambda})- \Gamma_- (p;\bm{\lambda})\right] p~,
	\\
	\dot \phi (t) &=& \omega (p;\bm{\lambda})~.
	\end{eqnarray}
\end{subequations}
Here, $\Gamma_+ (p;\bm{\lambda})$ is the damping and $\Gamma_- (p;\bm{\lambda})$ the anti-damping, resulting, typically, from injection of spin current. The frequency is given by $\omega (p;\bm{\lambda})$. The frequency and both damping and anti-damping depend on the power $p$. This dependence stems from non-linearities in the magnetization dynamics. The damping, anti-damping, and frequency do not depend on the phase for a circular spin-torque oscillator, which is what we consider here. They do, however, depend on a set of parameters ${\bm \lambda} = (\lambda_1, \cdots, \lambda_N)^T$, e.g., current and field, that are varied slowly and such that $\bm{\lambda} (t)$ makes a closed loop in the space of parameters. Completing this loop gives rise to a geometric contribution to the phase. By following the steps in the derivation of the main text for this general case, one finds  the geometric phase 
\begin{equation}
\label{eq:geophasevectorpot}
\phi_{\rm geo} =  \oint d\bm{\lambda}\cdot {\bf A}~, 
\end{equation}
with the vector potential
\begin{equation}
\label{eq:vectorpot}
A_i = \left(\left. \frac{\partial \omega }{\partial p} \right|_{p_0} \right)  \left(\left. \frac{\partial  F }{\partial p} \right|_{p_0}\right)^{-1} \frac{\partial  p_0 }{\partial \lambda_i}~,
\end{equation} 
with $i \in \{1,\cdots, N\}$. Here,
\begin{equation}
F (p) = -2  \left[ \Gamma_+ (p;\bm{\lambda})- \Gamma_- (p;\bm{\lambda})\right] p~,
\end{equation}
and $p_0 = p_0 (\bm{\lambda})$ is determined by solving for $p$ in $F (p)=0$.
\subsection{Ellipticity}
In the case of a spin-torque oscillator with ellipticity, resulting, for example, from magnetic anisotropies that favor a certain direction for tilting of the magnetization away from the easy axis, the equations of motion become
\begin{subequations}
	\label{eq:EOMphiandpelliptical}
	\begin{eqnarray}
	\dot p(t) &=& -2  \left[ \Gamma_+ (p,\phi;\bm{\lambda})- \Gamma_- (p,\phi;\bm{\lambda})\right] p~,
	\\
	\dot \phi (t) &=& \omega (p,\phi;\bm{\lambda})~.
	\end{eqnarray}
\end{subequations}
As was shown in Ref.~\cite{PhysRevLett.66.847}, systems described by these equations can exhibit geometric phases that are different in origin than the one that arises in the circular case. It is hard to give a simple analytical expression for these geometric phases, and we do not consider them further. We do, however, note that this also implies that the phase difference between two coupled oscillators, which obeys an equation similar to that in Eq.~(\ref{eq:EOMphiandpelliptical}), may also exhibit a geometric phase shift in the regime where the phases are not locked.

\section{Arrays of spin-torque oscillators} To improve phase stability and output signal, one often considers arrays of spin-torque oscillators. Such arrays exhibit similar geometric phases, as we will discuss now. We consider $M$ coupled uni-axial spin-torque oscillators. The generic equations of motion are
\begin{subequations}
	\label{eq:EOMphiandparray}
	\begin{eqnarray}
	\dot p_\delta(t) &=& F_\delta ({\bf p}, \bm{\phi}; \bm{\lambda})~,
	\\
	\dot \phi_\delta (t) &=& \omega_\delta ({\bf p}, \bm{\phi};\bm{\lambda})~,
	\end{eqnarray}
\end{subequations} 
where $\delta=1, \ldots, M$ labels the power $p_\delta$ and phase $\phi_\delta$ of each oscillator, ${\bf p} = (p_1, \cdots, p_M)^T$, and $\bm{\phi} = (\phi_1, \cdots, \phi_M)^T$. Like before, the vector $\bm{\lambda}$ consists of $N$ system parameters that may be varied adiabatically. 

We rewrite these equations of motion in terms of the global phase $\phi=\sum_{\delta=1}^M \phi_\delta$ and $M-1$ phase differences $\Delta \phi_\nu=\Delta \phi_{\nu+1}-\Delta \phi_\nu$, with $\nu=1, \ldots, M-1$. This yields equations of the form
\begin{subequations}
	\label{eq:EOMphiandparrayrewritten1}
	\begin{eqnarray}
	\dot p_\delta(t) &=& F_\delta ({\bf p}, \bm{\Delta \phi}; \bm{\lambda})~,
	\\
	\Delta \dot \phi_\nu (t) &=& \Delta \omega_\nu ({\bf p}, \bm{\Delta \phi};\bm{\lambda})~, \\
	\dot \phi (t) &=& \omega ({\bf p}, \bm{\Delta \phi}; \bm{\lambda})~.
	\end{eqnarray}
\end{subequations} 
Crucially, spin-rotation symmetry around the $z$-direction ensures that the functions $F_\delta$, the frequency differences $\Delta \omega_\nu = \omega_{\nu+1}- \omega_\nu$, and the total frequency $ \omega = \sum_{\delta=1}^M \omega_\delta$ do not depend on the global phase itself.

We now consider an adiabatic excursion of the parameters $\bm{\lambda}$. For notational convenience, we introduce the vector ${\bf f} = (p_1, \cdots, p_M, \Delta \phi_1, \cdots, \Delta \phi_{M-1})$ that has $2M-1$ components, and rewrite the equations of motion to
\begin{subequations}
 	\label{eq:EOMphiandparrayrewritten2}
 	\begin{eqnarray}
 	\dot f_\mu &=& {\mathcal F}_\mu ({\bf f};\bm{\lambda})~, \\
 	\dot \phi (t) &=& \omega ({\bf f}; \bm{\lambda})~.
 	\end{eqnarray}
\end{subequations} 
We denote with ${\bf f}_0 (\bm{\lambda})$ the solutions of the $2M-1$ equations ${\mathcal F}_\mu ({\bf f}_0)=0$, where $\mu$ runs from $1$ to $2M-1$. In the adiabatic limit we have that 
\begin{equation}
{\bf f} = {\bf f}_0 + \left( \frac{\partial { \bm {\mathcal F}}}{\partial {\bf f}}\right)^{-1} \cdot \frac{\partial {\bf f}_0}{\partial t}~,
\end{equation}
where $\partial { \bm {\mathcal F}}/\partial {\bf f}$ is the matrix with elements $\partial {\mathcal F}_\mu/\partial f_{\mu'}$ on its $\mu$-th row and $\mu'$-th column, and the inverse in the above equation is a matrix inverse. Insertion in the equation for the global phase yields
\begin{equation}
\dot \phi (t) = \omega ({\bf f}_0;\lambda) + \frac{\partial \omega}{\partial {\bf f}} \cdot \left( \frac{\partial { \bm {\mathcal F}}}{\partial {\bf f}}\right)^{-1} \cdot \frac{\partial {\bf f}_0}{\partial t}~,
\end{equation}
from which one obtains the geometric phase as in Eq.~(\ref{eq:geophasevectorpot}),
with the vector potential
\begin{equation}
\label{eq:vectorpotarray}
A_i = \frac{\partial \omega}{\partial {\bf f}} \cdot \left( \frac{\partial { \bm {\mathcal F}}}{\partial {\bf f}}\right)^{-1} \cdot \frac{\partial {\bf f}_0}{\partial \lambda_i}~,
\end{equation} 
where $i \in \{1,\cdots, N\}$.

\end{document}